%% file: example_paper.tex
\documentclass[accepted]{article}
\usepackage[symbol]{footmisc}

\usepackage[nointegrals]{wasysym}
\usepackage{tikz}
\usepackage{enumitem}

\usepackage{booktabs}

\usepackage{filecontents}

\usepackage{microtype}
\usepackage{graphicx}
\usepackage{subfigure}
\usepackage{booktabs} 

\usepackage{hyperref}

\usepackage{icml2020}

\usepackage{filecontents}

\input{math_commands.tex}

\newcommand{\pr}{\text{Pr}}

\newcommand{\am}{\texttt{argmax}\ }
\newcommand{\lrn}{\texttt{LRN}\ }
\newcommand{\lrnf}{\texttt{LRN-Free}\ }

\newcommand*{\strain}{\ensuremath{D_\text{train}^\text{shadow}}}
\newcommand*{\stest}{\ensuremath{D_\text{test}^\text{shadow}}}
\newcommand*{\vtrain}{\ensuremath{D_\text{train}^\text{victim}}}

\newcommand*{\vtest}{\ensuremath{D_\text{test}^\text{victim}}}

\newcommand{\myparagraph}[1]{
\paragraph{\textbf{#1}}
}

\icmltitlerunning{Sampling Attacks: Amplification of Membership Inference Attacks by Repeated Queries}

\begin{document}
\twocolumn[
\icmnote{This is an extended version of our paper at the~\href{https://priml-workshop.github.io/priml2019/\#abs40}{PriML} workshop at NeurIPS'19.}
\icmltitle{Sampling Attacks: Amplification of Membership Inference Attacks by Repeated Queries}



\icmlsetsymbol{equal}{*}

\begin{icmlauthorlist}
\icmlauthor{Shadi Rahimian}{to}
\icmlauthor{Tribhuvanesh Orekondy}{goo}
\icmlauthor{Mario Fritz}{to}
\end{icmlauthorlist}

\icmlaffiliation{to}{CISPA Helmlotz Institute for Information Security, Saarbr{\"u}cken, Germany}
\icmlaffiliation{goo}{Max Planck Institute for Computer Science, Saarbr{\"u}cken, Germany}

\icmlcorrespondingauthor{Shadi Rahimian}{shadi.rahimian@cispa.saarland}
\icmlcorrespondingauthor{Tribhuvanesh Orekondy}{orekondy@mpi-inf.mpg.de}
\icmlcorrespondingauthor{Mario Fritz}{fritz@cispa.saarland}

\icmlkeywords{Machine Learning, ICML}

\vskip 0.3in
]



\printAffiliationsAndNotice{}  

\begin{abstract}
 Machine learning models have been shown to leak information violating the privacy of their training set. We focus on membership inference attacks on machine learning models which aim to determine whether a data point was used to train the victim model. \\
 Our work consists of two sides: We introduce sampling attack, a novel membership inference technique that unlike other standard membership adversaries is able to work under severe restriction of no access to scores of the victim model. We show that a victim model that only publishes the labels is still susceptible to sampling attacks and the adversary can recover up to $100\%$ of its performance compared to when posterior vectors are provided. \\
 The other sides of our work includes experimental results on two recent membership inference attack models and the defenses against them. For defense, we choose differential privacy in the form of gradient perturbation during the training of the victim model as well as output perturbation at prediction time. We carry out our experiments on a wide range of datasets which allows us to better analyze the interaction between adversaries, defense mechanism and datasets. We find out that our proposed fast and easy-to-implement output perturbation technique offers good privacy protection for membership inference attacks at little impact on utility.
\end{abstract}

\input{introduction}
\input{membership-inference}
\input{defenses}

\input{sampling}
\input{experiments}

\input{conclusion}

\bibliographystyle{icml2020}
\bibliography{references.bib}

\end{document}

%% file: math_commands.tex
\usepackage{amsmath,amsfonts,bm}

\def\eqref#1{equation~\ref{#1}}

\def\1{\bm{1}}

\DeclareMathAlphabet{\mathsfit}{\encodingdefault}{\sfdefault}{m}{sl}
\SetMathAlphabet{\mathsfit}{bold}{\encodingdefault}{\sfdefault}{bx}{n}

\def\gD{{\mathcal{D}}}

\def\gM{{\mathcal{M}}}

\def\gR{{\mathcal{R}}}

\DeclareMathOperator*{\argmax}{arg\,max}

%% file: introduction.tex
\section{Introduction}
\label{sec:introduction}

Recent machine learning (ML) methods -- especially deep learning approaches -- largely owe their success to the availability of big data and the computation power to train huge models with millions of parameters on this \textit{training dataset}. Often, people might assume that since these models are designed to learn statistical properties of their training dataset, they protect the privacy of the individuals who contribute to these datasets. Unfortunately, this is not the case and the ML models are proven to violate the privacy of this training set in many unintended ways such as memorizing details about the individuals~\cite{song2017machine}, leaking features of their training set~\cite{melis2019exploiting} and contributing to re-identification problems e.g. via saving users' geodata~\cite{maouche2017ap}. 

In this study we are concerned with membership inference (MI) attacks. These are attacks carried out by an adversary who has complete or partial data records and access to a trained victim ML model and wishes to know whether that data record has been used to train the victim model. MI attacks pose serious threats on the privacy of the individuals present in databases and have been successfully applied on models trained on image data, medical data, transaction records, etc.~\cite{nasr2019comprehensive,naranyanan2008robust, shokri2017membership, song2019privacy}. 

MI attacks are carried out in a black-box setting, where adversary has no access to the internal parameters of the victim model. The assumption is that posterior vectors from the victim classifier suffice to reveal the membership status of queried points. For our experiments, we chose two standard MI adversaries: One that learns to distinguish members of the training set from non-members and one with prior belief on their differences~\cite{salem18}.  

We employ differential privacy (DP)~\cite{dwork2014algorithmic} to defend against MI attacks. DP is becoming more and more popular as a method to protect the privacy of individuals involved in a machine learning setting. It offers flexible techniques which can be applied locally on the data~\cite{kairouz2014extremal,qin2016heavy}, or during the training on training parameters such as gradients~\cite{abadi2016deep} or the objective function~\cite{zhang2012functional} or it can be applied on the output of the function~\cite{mcsherry2007mechanism, rastogi2009relationship}. We select DP-Stocahstic Gradient Descent (DP-SGD)~\cite{abadi2016deep} which is an extensively-used gradient perturbation technique and compare the protection it offers against MI attacks to an output perturbation technique. 

We also consider the extreme case of protection when the scores are not published by the victim model and the adversary only receives the predicted labels. This results in standard MI attack methods to fail to perform. However, we introduce a new \textit{sampling} attack model which is constructed to utilize only the labels. We show how in practice this novel method compares with standard MI adversaries.   

\myparagraph{Our contributions:}
\begin{itemize}
\item We introduce our novel \textit{sampling attack} model which performs membership inference under the severe restriction of no access to confidence scores of the attacked classifier. 
\item We compile a comprehensive list of all the prominent datasets in membership attack studies and compare them under unified metrics and alongside each other. This helps us better understand and analyze membership inference attacks. 
\item Focusing mainly on the practical implications of differential privacy (DP)~\cite{dwork2014algorithmic} rather than the theoretical bounds on the privacy budget, we use DP as a method to defend against membership inference attacks. We show the interplay between datasets, attack models and defenses. 

\item Considering the practical limitations of applying DP during the training of the models, we formulate an output perturbation method that can be applied on models at prediction time. We also study the effect of this method, which we call \textit{DP-Logits}, as a defense for membership attacks and contrast it with DP-SGD~\cite{abadi2016deep}.
\end{itemize}

The structure of this paper is as follows: We first introduce membership inference attacks in details in Section~\ref{sec:membershipinference}. We then summarize possible defense methods against membership adversaries in Section~\ref{sec:defenses}, with a focus on differential privacy as a defense in Section~\ref{sec:dp}. We formulate the novel sampling attack and explain the details and practical methods to implement it in Section~\ref{sec:sampling}. We finally presents our experimental results in Section~\ref{sec:experiments}. 

%% file: membership-inference.tex
\section{Membership Inference}
\label{sec:membershipinference}
Membership inference attack decides whether or not a certain data point is a member of a dataset. The privacy risks of membership inference attacks were first brought into attention by Homer et al.~\cite{homer2008resolving} when they demonstrated that they could successfully resolve the presence of an individual in a highly-complex DNA mixture. One of their key findings is that publishing only the composite statistics over a collection of genomic data would not protect the privacy of the individuals who are members of that collection. In a follow up paper~\cite{wang2009learning} they use more sophisticated test statistics to achieve better results with less prior knowledge on the victims. Similar attacks on other biomarkers such as microRNA have also been successfully performed~\cite{backes2016membership}.

\subsection{Membership Inference on Machine Learning Models}
\label{sec:mi-in-ml}
The increased popularity of machine learning models translates to an ever-increasing need for data to train these models. This data often contains sensitive information from individuals and protecting the privacy of it is of great importance. Therefore, our focus in this paper is on membership inference attacks on machine learning models.

 These models are usually trained on a set of data points $x$ so that the function $f(x; \theta)$, which is characterized by parameters $\theta$, is learned. In this context, the goal of the adversary is to determine: given a trained \textit{victim} model, is a certain data point $x_i$ a member of the training set of this model? i.e:
\begin{equation}
    A(x;\phi): X\rightarrow \{0\ \text{(non-member)},1\ \text{(member)}\}
\end{equation}
where $A$ is the adversary, $\phi$ are the parameters that the adversary utilizes and $X$ is the space of possible data points. 

Now we will explain two adversarial models that cover standard membership attacks. The first adversary requires training with the objective of finding statistical differences between members and non-member data points; whereas the second model has a prior belief on these statistical differences and no learning phase is required.  

\myparagraph{Learning Based Adversary} 
First introduced in~\cite{salem18}, this adversary relies on training a \textit{shadow} model and a binary \textit{attack} model. Shadow models mimic the behavior of the victim model and are in the possession of the adversary. The adversary can freely study the behavior of these models as a surrogate for the victim models with restricted access. The task of the binary attack classifier is to classify its input as member/non-member of the training set. 

The following steps are taken by the adversary:
\begin{enumerate}
    \item \textbf{Shadow model training:} Train the shadow model with a set of data point from the same distribution as the training set of the vicitm model.
    \item \textbf{Binary classifier training:} Query the trained shadow model with its training set as well as a hold-out test set. Collect the posteriors and pass them to the binary attack classifier. 
    \item \textbf{Attack the victim model:} Query the victim model with the desired data points and use the trained binary classifier to decide the membership status. 
\end{enumerate}


The assumption about this attack is that the adversary has a black box access to the victim model and can only study the returned posterior vectors. In this method, only one shadow model and one binary classifier are used. This can be viewed as a relaxed version of multiple shadow models and multiple binary classifiers of Shokri et al.~\cite{shokri2017membership}. 

We will refer to this attack model as \lrn adversary.

\myparagraph{Learning Free Adversary} The dependence of the \lrn adversary on the shadow model and the binary classifier as well as data to train these models, is an undesirable factor. For this reason, Salem et al.~\cite{salem18} suggest a more versatile model that requires no shadow model or attack binary classifier. A black-box access to the victim model is also assumed for this attack. The posterior vectors from the victim model are inspected and if the maximum element exceeds a calibrated threshold it will be classified as a member, otherwise non-member. This can be summarized in the following function:
\begin{equation}
    A(x; T) = 
     \begin{cases}
       1 \quad\text{if}\quad \max_y \pr(y|x)\geq T \\
       0\quad\text{otherwise}\\
     \end{cases}
\end{equation}
which is parameterized by a single threshold $T$. The adversary queries the victim model by data point $x$ and decides based on the maximum of the returned posterior vector $\pr(y|x)$ whether or not it is a member of the training set of the vicitm model.

The name "learning free" comes from the fact that no shadow or attack classifiers are trained. We will refer to this attack as \lrnf adversary. 

%% file: defenses.tex
\section{Defenses for Membership Inference Attacks}
\label{sec:defenses}
In the previous section we defined the membership inference attacks and described two generic models to carry out these attacks in practice. Now, we will explore methods to defend against them. 

To defend against these attacks, we need to understand what factors make these attacks possible and how we can limit and paralyze the adversary. Most of the previous work on defenses against membership inference attacks can be summarized into two categories:

\myparagraph{Generalization-based techniques}
~\cite{shokri2017membership} was the first to define the membership inference attacks in a machine learning setting. They also identify the overfitting of the victim model as one of the main culprits for vulnerability to membership inference attacks. They hypothesize that the victim model memorizes its training set such that the posteriors show a statistical difference between the seen and hold-out data. A more comprehensive study about the correlation of overfitting to membership inference attacks can be found in~\cite{yeom2018privacy}. 

These findings prompt a line of defense that relies on enforcing generalization on the victim model. ~\cite{shokri2017membership} suggest using L2 regularization of the parameters and restricting the number of training epochs. ~\cite{salem18} use dropout and ensemble learning to train the victim model to help it generalize better. In a slightly different approach, ~\cite{nasr2018machine} utilizes adversarial training of the victim model in the form of a min-max game to help the model generate indistinguishable predictions on its training set and an unseen dataset.

\myparagraph{Noising-based techniques} Adding randomness to different parameters of the victim model at different stages is one of the most natural ways to confuse any adversary. In fact, the first defenses against membership inference attacks on the genome data~\cite{wang2009learning} proposes adding carefully-crafted noise to the published dataset. 

Jia et al~\cite{jia2019memguard} suggest adding noise to the output of the victim model. They generate specially-composed noise vectors for the victim model's posteriors such that they act as adversarial examples for the attacker. 

In a more formal and mathematics-driven line of work \textit{differential privacy} is leveraged to add noise the gradients during the training of the victim model~\cite{jayaraman2019evaluating, rahman2018membership}. 

In this work we mainly focus on differential privacy as a defense since it is a well-defined privacy framework and very flexible with respect to the methods that can be applied to build a differentially-private model. In the rest of this section we introduce differential privacy and explain how it can be utilized in our setting. 

\subsection{Differential Privacy}
\label{sec:dp}
Differential privacy (DP)~\cite{dwork2014algorithmic, dwork2015robust, dwork2011firm, dwork2006calibrating} is a mathematical definition bounding the maximum divergence between the probability distributions of the outputs of a mechanism $M$ when it is applied on two \textit{adjacent datsets} $d$ and $d'$. Two datasets $d, d'\in \mathcal{D}$ are adjacent when they differ in only one entry, e.g. when the data of one user is removed from one of the two identical datasets. 

\myparagraph{Definition} We formally define a differentially private algorithm $M: \gD \rightarrow \gR$ when the following condition holds: 
\begin{equation}
	\label{eq:dp:main}
	\pr[M(d)\in S]\leq e^{\epsilon}\ \pr[M(d')\in S] + \delta\\
\end{equation}
where $M$ is a randomized algorithm with domain $\gD$ of all possible datasets and range $\gR$, $d$ and $d'$ are two adjacent datasets and $S\subseteq \gR$ is the output of the algorithm $M$. The privacy parameters (privacy budget) $\epsilon$ and $\delta$ bound the probability of the output being more likely for one dataset compared to the other. 

We say that the randomized algorithm $M$ is $(\epsilon, \delta)$-differentially private if Equation.~\ref{eq:dp:main} holds for some parameters $\epsilon, \delta\geq0$. It is usually suitable to set the value of $\delta=1/|d|$ where $|d|$ is the number of data points in the dataset. So we are mostly concerned with and aim to achieve a smaller value of $\epsilon$ since it guarantees more privacy.

\myparagraph{Properties of differential privacy} Next, we will introduce two of the most important properties of differential privacy: 
\begin{enumerate}
    \item Immunity to post processing~\cite{dwork2014algorithmic}: Any function of the output of the $(\epsilon, \delta)$-DP algorithm $M$ is $(\epsilon, \delta)$-DP. This means that an adversary without any additional knowledge on the dataset, is unable to extract more information about the dataset with further analysis of $M$. 
    \item Composition~\cite{dwork2014algorithmic}: The combination of two or more DP algorithms is also a DP algorithm with degraded privacy budgets $\epsilon'$ and $\delta'$. There are various theorems which act under certain conditions to calculate $\epsilon'$ and $\delta'$ given the privacy budgets of each initial mechanism. In the simplest form, the independent use of $k$ mechanisms, each $(\epsilon, 0)$-DP is $(k\epsilon, 0)$-DP.
    \end{enumerate}
    
\myparagraph{Some DP Mechanisms} Now that we are familiar with the basics of differential privacy we can look at some mechanisms that ensure DP:
\begin{enumerate}
    \item Randomized response mechanism~\cite{warner1965randomized}:
    One of the oldest privacy-ensuring mechanisms, randomized response was first used to protect participants in surveys with a "yes/no" possible answer. 
Assume we are interested whether a property $P$ is present in participants. Participants are asked to flip a coin then:

    \textit{if tails:} respond truthfully

    \textit{if heads:} flip a coin again. If \textit{heads} answer "yes" otherwise "no"
    
For this mechanism $\epsilon$ can be calculated as:
\begin{align}
    e^\epsilon = \frac{\pr(\text{answer}=\text{yes}|\text{truth}=\text{yes})}{\pr(\text{answer}=\text{yes}|\text{truth}=\text{no})} =\frac{3/4}{1/4}=3\nonumber
\end{align}
So the randomized response mechanism is $(\ln(3), 0)$ differentially private.

    \item Gaussian Mechanism~\cite{dwork2014algorithmic}:
    Consider a deterministic real-valued function $f: \gD\rightarrow \mathbb{R}$, we define the \textit{$l_2$-sensitivity} of $f$ as:
\begin{equation}
	\label{eq:dp:sensitivity}
	\Delta_2(f) = \max_{d, d'\ \text{adjacent}} \|f(d)- f(d')\|
\end{equation}
We can approximate $f$ with an $(\epsilon, \delta)$-DP function $M$ by adding Gaussian noise to it:
\begin{equation}
	\label{eq:dp:gauss}
	\gM(d) = f(d) + \mathcal{N}(0, \sigma^2)
\end{equation}
where $\mathcal{N}(0, \sigma^2)$ is a Gaussian distribution with mean 0 and standard deviation $\sigma \geq {c \Delta_2(f)}/{\epsilon}$, and 
$c^2>2\ln({1.25}/{\delta})$.
    
\end{enumerate}

\subsubsection{Differential Privacy for Machine Learning}
\label{sec:dp-ml}
For sophisticated processes which require a combination of multiple functions, we can utilize and combine DP mechanism and use the composition theorems to keep track of the accumulated privacy budget. In this paper we work with deep learning models which are non-linear combination of functions that are trained over multiple iterations. If we aim to make a deep learning model differentially private, we need to be cautious about the DP methods that are applied and how to correctly track the privacy budget. In this section, we introduce two possible DP methods for deep learning: 
\myparagraph{DP-Stochastic Gradient Descent (DP-SGD)~\cite{abadi2016deep}} This method achieves privacy by adding noise to the parameters of the model during the training. First, the gradients are clipped then an additive noise is applied to them:
\begin{align}
\bar{\mathbf{g}}_t(x_i)\gets 
		\mathbf{g}_t(x_i)/\max(1, \frac{\|\mathbf{g}_t(x_i)\|_2}{C})\\
		\tilde{\mathbf{g}}_t\gets\frac{1}{L}
		(\sum_i\bar{\mathbf{g}}_t(x_i)+\mathcal{N}(0, (\sigma C\mathbb{I})^2))
\end{align}
where $\textbf{g}_t$ is the gradient vector at epoch $t$, $C$ is the clipping norm, $L$ is the number of samples randomly chosen for calculation of the gradient and $\mathcal{N}(0,\sigma^2)$ is a Gaussian with standard deviation $\sigma$. This process can be viewed as Gaussian mechanism on top of the stochastic gradient descent. The reason behind the clipping step is to bound the $l2$-sensitivity of the gradients. 

The final privacy budget after the training can be calculated with \textit{moments accountant} which is a composition method for a sequence of adaptive mechanisms. 

\myparagraph{DP-Logits} 
While DP-SGD provides a strong guarantee, for practical black-box deployment scenarios, these guarantees significantly hinder performance. For example, applying DP-SGD requires clipping of the gradients in each step of the training which results in a slow-downed training process. Also the accumulated value of $\epsilon$ can potentially grow to very large numbers for complex neural network architectures that require many iterations of training and are sensitive to noise. Furthermore, the party in the possession of a model may not always have access to the training phase of the model. Consequently, we now work towards an approach that perturbs the model outputs (rather that \textit{all} parameters) to specifically safeguard against black-box membership inference attacks. This method can be rapidly applied on top of any trained model.

For this, we use Gaussian mechanism to add noise to $l2$ normalized logits at prediction time: 

\begin{align}
&\bar{\mathbf{l}}(x_i) \leftarrow \mathbf{l}(x_i) /\max(1, \frac{\|\mathbf{l}(x_i)\|}{S})\\
&\tilde{\mathbf{l}}(x_i)\leftarrow \bar{\mathbf{l}}(x_i)+\mathcal{N}(0, (\sigma S \mathbb{I})^2)
\end{align}

where $\mathbf{l}(x_i)$ indicates the vector of logits for the input $x_i$ and $S$ is the clipping norm. The privacy budget with the noise scale $\sigma$ and $l_2$ sensitivity $S$ can be calculated as:
\begin{align}
    \label{eq:dplogit}
  \sigma \geq \frac{S q}{\epsilon}\sqrt{2\ln(\frac{1.25}{\delta})}\qquad \text{for } \delta=\frac{1}{|d|}  
\end{align}
where $|d|$ is the size of the training dataset, and $q$ the number of queries. As opposed to the DP-SGD method which is applied during the training and leaves the trained model immune to post-processing, adding noise to the logits at the prediction time suffers from degradation by the number of queries. Especially important is to restrict the number of times that a same point can be queried, as an adversary would be able to zero out the noise by repeated querying. However, this method has the advantage of being fast and easy to implement. 

\subsection{Argmax Defense}
\label{sec:argmax}
Diverging from differential privacy and the other suggested defense methods, we can take a step back and tackle this problem from a different perspective. To date, all the membership inference adversaries that we are aware of, rely on and utilize the posterior vectors from the victim model. This means that if the victim model returns the most confident `argmax' label $k=\argmax_k\pr(y=k|x_i)$ instead of the full posterior, the adversary is unable to carry out the attack. We refer to this method as \am defense. 

Note that the \am defense is not always feasible, mostly as a result of the problem design setting where the scores are required and expected by the benign user.

%% file: sampling.tex
\section{Sampling Attack}
\label{sec:sampling}

In Section~\ref{sec:argmax} we argued that the \am defense is effective against all the previously-suggested membership inference adversaries due to their dependence on the posteriors vectors. Now we will introduce our novel attack method which is designed to work under this severe restriction. 

Attacks under limited access to the posterior vectors have been studied before, for example,~\cite{dang2017evading} propose evasion attacks when the score of the detector
is not accessible to the adversary. To fool the detector, they generate perturbations of the malicious sample and attempt to find the first perturbed sample that traverses the malicious/benign boundary of the detector. However, their attack does not directly depend on the posterior vectors and the goal is to only evade the detector.

Membership inference adversaries are designed to study the posterior vector, so the idea behind our sampling adversary is to reconstruct these vectors from the returned labels. We achieve this by populating a sphere around each data point with multiple perturbations of it and counting the number of perturbed samples that fall under each label. 

If $x'$ is a perturbation of data point $x$, from probability theory we have: 

{\small
\begin{align}
    \pr(y=c|x=x_i) &= \sum_{x'\in\mathcal{X}'_i} \pr(y=c, x'|x=x_i)\label{1line}\\
    &= \sum_{x'\in\mathcal{X}'_i} \pr(y=c| x', x=x_i)\pr(x'|x=x_i)\label{2line}\\
     &= \sum_{x'\in\mathcal{X}'_i} \pr(y=c| x')\pr(x'|x=x_i)\label{3line}
\end{align}
}%
where $\mathcal{X}'_i$ is the space of possible perturbations of $x_i$. Line~\ref{1line} is the marginalization over the perturbed data points $x'$ and line~\ref{2line} is the expansion according to the Bayes' theorem. The transition to line~\ref{3line} comes from the fact that the victim models decision on the posterior of $x'$ does not depend on the unperturbed point $x_i$. The probability $\pr(x'|x)$ is the perturbation function acting on $x$. So this can be formulated as a Monte-Carlo integration~\cite{mackay1998introduction} in the form of: 
{\small
\begin{align}
    \label{eq:montecarlo}
    \pr(y=c|x=x_i) &= \frac{1}{N} \sum_{x'} \pr(y=c| x')\quad \text{where } x'\sim pert(x_i)
\end{align}
}%

But we do not have access to the posteriors $\pr(y|x')$, so we propose relaxing Eq.~\ref{eq:montecarlo} and using the returned labels instead. Algorithm.~\ref{algo:sampling} demonstrates how the sampling adversary works. The perturbation function pert() acts on each data point $x$ to generate $N$ perturbed samples and the identity function $\mathbf{I}$ builds histograms over the labels of the perturbed points. We hypothesize that this histogram can be a suitable replacement for the posterior. 

\begin{algorithm}
	\caption{Sampling Attack}
	\label{algo:sampling}
	\begin{algorithmic}
	\STATE {\bfseries Input:}
	Data points $\{x_1,\ldots,x_M\}$, perturbation function pert$(x_i;p)$. Parameters: number of perturbations $N$, perturbation scale $p$.\\
	\FOR{$i\in [M]$}
	\FOR{$n\in [N]$}
	\STATE get labels: $l= \text{pert}(x_i; p)$
	\ENDFOR
	\STATE build histograms: $\pr(y=c|x_i) = \frac{1}{N}\sum \mathbf{I}(l=c)$
	\ENDFOR
	\STATE {\bfseries Output:} Posterior vectors $\pr(y|x)$
	\end{algorithmic}
\end{algorithm}

In practice, the following steps are taken by the adversary:

\begin{enumerate}
    \item \textbf{Shadow model training:} Train a shadow model with data from the same distribution as the training set of the victim model.
    \item \textbf{Sampling on the shadow model:} Execute algorithm.~\ref{algo:sampling} for the training set as well as a hold-out test set of the shadow model. Repeat for different perturbation levels $p$.
    \item \textbf{Attack the shadow model:} Using the reconstructed posteriors from the previous step, attack the shadow model with one of the conventional adversaries.
    \item \textbf{Attack the victim model:} Choose the optimum value of $p$ according to some performance metric of the adversary. Attack the victim model with the chosen $p$ value and the adversary from step 3. 
\end{enumerate}

In Table.~\ref{tab:attacks} we compare the sampling attack with the \lrn and \lrnf adversaries. The sampling attack requires the training of a shadow model. If we choose \lrnf adversary for steps 3 and 4 of the sampling attack, no binary classifier training is required.  
\begin{table*}[!htbp]
    \centering
      \caption{Comparison of \lrn and \lrnf to the sampling adversary. Full circles mean that the condition is required and the half full circle means flexibility in terms of training.}
    \begin{tabular}{cccccc}
    \toprule 
        adversary&shadow model&binary classifier&data distribution access&posterior access&training\\
        \midrule
         \lrn& \newmoon&\newmoon&\newmoon&\newmoon&\newmoon \\
         \lrnf& -&-&-&\newmoon&-\ \\
         sampling&\newmoon&-&\newmoon&-&\LEFTCIRCLE\\
         \bottomrule
    \end{tabular}
    \label{tab:attacks}
\end{table*}

%% file: experiments.tex
\section{Experiments}
\label{sec:experiments}
In this section we present the experimental results of membership inference attacks and the effectiveness of the suggested defenses against them.

For the first two parts we mostly focus on the interplay of adversaries and defense mechanisms from a practical point of view. To this end we choose a collection of 8 different datasets that were used in the most influential membership inference attack studies (e.g.~\cite{shokri2017membership, salem18, jia2019memguard, nasr2018machine}) to be able to compare the results with one unified measure. 

We choose two differentially-private defenses (see Section.~\ref{sec:dp}), one that perturbs the parameters of the model and one that perturbs the outputs, and apply them on the adversaries. 

At last, with an insight into how adversaries work and what datasets are worth further investigating, we present the results on our novel sampling attack technique.

\subsection{Datasets}
\label{sec:datasets}
The datasets that we begin our experiments with, are:

\paragraph{MNIST}The MNIST\footnote{\label{foot:mnist}%
		http://yann.lecun.com/exdb/mnist/} dataset consists of $60,000$ training data and $10,000$ test data of grayscale images of size $28\times 28$. These images depict handwritten digits $(0-9)$ and are centered with respect to the frame of the image.

\paragraph{FashionMNIST}%
Created by Zalando\footnote{\label{foot:fmnist}%
		https://www.kaggle.com/zalando-research/fashionmnist}, this dataset consists of $60,000$ and $10,000$ training and test set data points, respectively. These images are also $28\times28$ and in grayscale and represent 10 classes of fashion items such as "tops", "trousers", "sneakers", etc

\paragraph{CH-MNIST}
This preprocessed dataset obtained from Kaggle\footnote{\label{foot:chmnist}%
		https://www.kaggle.com/kmader/colorectal-histology-mnist} contains $5000$ greyscale images of different types of tissue in colorectal cancer patients. The size of images is $64\times64$ and the task is to classify these images into one the 8 possible tissue categories.

\paragraph{CIFAR10, CIFAR100} We used CIFAR-10 and CIFAR-100\footnote{\label{foot:cifar}%
		https://www.cs.toronto.edu/~kriz/cifar.html} for our experiments. Both consists of color images of size $32 \times 32$ and have $50,000$ raining data and $10,000$ test data. CIFAR-10 has 10 classes such as "air plane", "dogs", "cats", etc.:5000 randomly-selected images per class in its training set and 1000 randomly-selected images per class in its test set. On the other hand, CIFAR-100 has 20 super classes, each containing 5 class (in total 100 classes) of different subjects such as animals, humans and vehicles. Similar to CIFAR-10, it also has 5000 randomly-selected images per class in its training set and 1000 randomly-selected images per class in its test set.

\paragraph{Purchase100}
Purchase\footnote{\label{foot:purchase}%
		https://www.kaggle.com/c/acquire-valued-shoppers-challenge/data} is dataset of shopping history of several thousand customers and the aim is to classify the customers into $k$ different classes of shopping styles so that accurate coupon promotions can be suggested to them. This dataset has no ground truth for the labels. Similar to~\cite{salem18, shokri2017membership}, we use a simplified version of this dataset with $\sim 200,000$ data points and a 600-dimensional vector of purchases per data point where each element can take a value of either 0 or 1 (present or not present in the shopping history). Afterwards, k-means clustering algorithm~\cite{lloyd1982least} is used to cluster these vectors into 100 classes. We call this version of the Purchase dataset with 100 classes Purchase100. 

\paragraph{Texas100} This includes patients' data published by the Texas Department of State Health Services\footnote{\label{foot:texas}%
		https://www.dshs.texas.gov/THCIC/Hospitals/Download.shtm}. This dataset contains $6,169$ binary features of $67,330$ patients, such as diagnosis of various disease, procedures performed on the patient and other properties of each patient. We use a preprocessed version obtained from~\cite{jia2019memguard}. Given the input data of each patient, the task is to choose among the most suitable procedure among the available 100 most frequent ones.

\paragraph{Location}
This dataset is the binary representation of $446$ locations\footnote{\label{foot:texas}%
		https://sites.google.com/site/yangdingqi/home/foursquare-dataset} (either visited or not visited) by users. This comes with $5,010$ data points and the task is to classify the datapoint into one of the 30 possible classes. We obtained a preprocessed version from~\cite{jia2019memguard}.

\subsection{Membership Inference Attacks on All Datasets}
\label{sec:mi-general}

Throughout our studies we encountered many papers on membership inference attacks and/or defenses against them which each used different datasets and different metrics to evaluate the success of the attacks and defenses. This motivated us to first combine all these datasets under a unified metric to be able to compare and better understand the results. We chose \lrn and \lrnf (see Section~\ref{sec:membershipinference}) as our conventional adversaries. Below we explain the details of our experiments:

\myparagraph{Evaluation metrics} To evaluate the performance of the adversary we chose Area Under the ROC Curve (AUC) since it is independent of the threshold that the adversary chooses to distinguish members from non-members and gives a better overview of the performance of the attacker. An AUC value of 0.5 means random guessing and implies completely unsuccessful attack, whereas AUC value of 1.0 implies a perfect attack. 

\myparagraph{Data splits} We divide each dataset into 4 equal parts \vtrain, \vtest, \strain and \stest. For this purpose, all of the training and test sets of MNIST, FashionMNIST, CH-MNINST, CIFAR10, CIFAR100 and Location datasets are combined and used. For Purchase100 and Texas100 we use a total of 80,000 and 40,000 randomly selected points, respectively.

\myparagraph{\lrn adversary}  We use \strain to train the shadow model. After training we query the shadow model by its training set and the unseen data points of \stest and use these posteriors to train the binary classifier. We then test the performance of the binary classifier on the posterior prediction of the victim model when queried by its training set \vtrain and the unseen set \vtest. To avoid choosing a decision threshold, at this stage we only take the output of the sigmoid function of the binary classifier. This allows us to calculate the AUC values over decisions of the binary classifier. 

\myparagraph{\lrnf adversary}
 For a fair comparison with the \lrn adversary we only use \vtrain and \vtest for our learning-free method. We query the trained victim model by its training data \vtrain and the hold-out \vtest and take the maximum element of the returned posterior vectors. Similar to \lrn adversary, we refrain from choosing a decision threshold and calculate AUC values over all the possible thresholds.

\begin{figure*}
    \centering
    \includegraphics[width=0.9\textwidth]{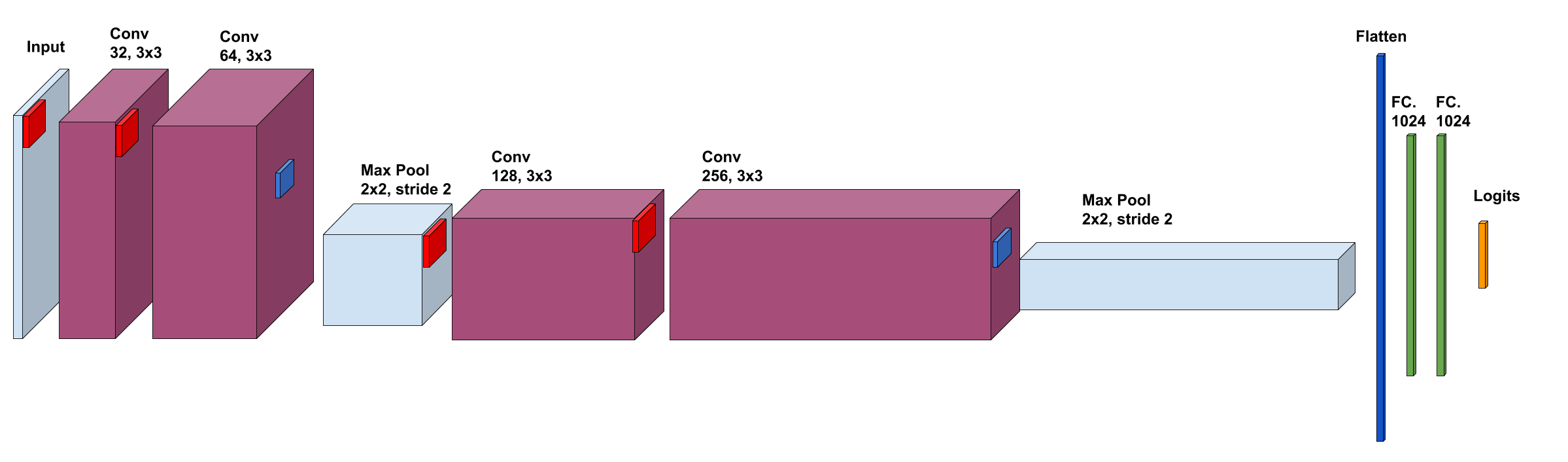}
    \caption{The structure of the CNN used for all the image datasets. All the red squares indicate $3\times 3$ convolution 
	filters and the blue squares show $2\times2$ pooling with stride of 2. In the end there are two fully-connected (FC) layers.}
    \label{fig:CNN}
\end{figure*}

\myparagraph{Victim model}
 For all the image datasets (MNIST, FashionMNIST, CH-MNIST, CIFAR10 and CIFAR100) we use a VGG-like~\cite{vgg} convolutional neural network (CNN) as shown in Figure.~\ref{fig:CNN}. For Location we use a fully connected neural network with layer sizes $[256, 128, 128, 30]$. For Texas 100 and Purchase100 we use a fully connected neural network with layer sized $[512, 256, 128, 100]$. 
We train these models with \texttt{AdamOptimizer} with learning rate of $0.001$. The maximum number of epochs is set to 50 but an early stopping criterion is also set.

\myparagraph{Shadow model}  We use the same structures as the victim model for the shadow model and train the model with \strain using the same procedure as the victim model. 

\myparagraph{Attack binary classifier}
 For the attack classifier we use a
neural network with one 64-unit hidden layer and a sigmoid output for the final binary classification. 

\myparagraph{Results and discussion} Table.~\ref{tab:acc} demonstrates the evaluated performance of the \lrn and \lrnf adversaries on each dataset. We  observe that the adversaries are more successful (higher AUC) on datasets with lower test accuracy. This can be contributed to the inability of the victim model to generalize well. Models that achieve higher test accuracy do not overfit on their training data and output posterior vectors that are statistically homogeneous across all the data. 

We can also see that the number of classes in each dataset plays a role in the success rate of adversaries. For datasets with comparable number of training points, those that have fewer number of classes are less prone to attacks. This can also be explained with the fact that it is easier for models to generalize and perform for fewer number of classes. 

We noticed that some datasets that were well-protected in our experiments, that is MNIST, FashionMNIST and CH-MNIST, were reported with high success rate of the adversary in the previous literature. We relate this to a better network structure and the application of an early stopping mechanism in the training such that overfitting is avoided. 

\begin{table}
\centering
\caption{Test accuracies versus performance of the \lrn and \lrnf attacker for all of the datasets.}
{\small
\begin{tabular}{lccccc}
    \toprule
     $\#$classes/dataset&$\text{accuracy}_{D^{victim}_{test}}$  & $\text{AUC}_{\text{LRN}}$& $\text{AUC}_{\text{LRN-Free}}$\\
     \midrule
     10 MNIST&0.98&0.50&0.51\\ 
     10 FashionMNIST&0.88&0.505&0.505\\
     10 CH-MNIST&0.76&0.52&0.505\\
     10 CIFAR10&0.69&0.60&0.59\\
     100 CIFAR100&0.35&0.76&0.75\\
     100 Texas100&0.56&0.72&0.67\\
     100 Purchase100&0.78&0.60&0.58\\
     30 Location&0.61&0.81&0.88\\
     \bottomrule
\end{tabular}
}%

\label{tab:acc}
\end{table}

\subsection{A Study of the Conventional Adversaries and Defenses for Them}
\label{sec:mi-defenses}
Now we will study the effect of differentially-private defenses on \lrn and \lrnf adversaries. We proceed with only CIFAR10/100, Texas100, Purchase100 and Location since the AUC value for both adversaries for the remaining datasets was close to chance level. 

For defense, we choose DP-SGD  that perturbs the parameters of the model during the training and DP-Logits that adds noise to the outputs of the model after training (see Section~\ref{sec:dp}). 

The adversarial models and their structure, the data split and the  evaluation metric are the same as the previous section. 

\myparagraph{DP-SGD} Clipping norm $C$ should be selected based on the distribution over the unclipped gradients for each dataset. We found that a value of $1\leq C\leq3$ works best for our datasets. Instead of noise level $\sigma$, we choose to report the noise multiplier $m=\sigma / C$ since the amount of noise needed to achieve a certain level of privacy is dependent on the clipping value $C$. We choose $m\in\{5 \times 10^{-4}, 10^{-3}, 5\times 10 ^{-3}, 0.01, 0.05, 0.1, 0.5, 1.0\}$ for the DP-SGD experiments.
The value of $\delta$ is set to $1/|D^{victim}_{train}|$ and $\epsilon$ is calculated by the accountants method.

\myparagraph{DP-Logits} To choose a clipping value $S$, for each dataset we calculated the histograms over the $l2$ norm of the unclipped logits. We then chose the 60th-percentile of these histograms as the value for $S$. Similar to DP-SGD we report in terms of noise multiplier $m=\sigma/S$. We chose $m\in\{10^{-5},5\times10^{-5},10^{-4}, 5\times 10^{-4},5\times10^{-3},10^{-3},5\times10^{-2},10^{-2}\}$ as the noise levels of this defense mechanism. We set the value of $\delta$ to $1/|D^{victim}_{train}|$ and according to Eq.\ref{eq:dplogit} $\epsilon$ can be calculated as: $$\epsilon = \frac{q}{m}\sqrt{2\ln(1.25*|D^{victim}_{train}|_1)}$$ 

For \lrnf adversary we set $q=1$ since the membership/non-membership status of each point is evaluated individually and independent of the results for other data points. 

For \lrn adversary we need to set $q=|D^{victim}_{train}+D^{victim}_{test}|$ since this is the number of times that the victim model is queried by an adversary that is learning based. 

\begin{table*}[!htbp]
\centering
    \hspace{-0.5cm}
     \caption{Optimal values of noise multipliers and corresponding privacy budgets, for both methods}
    \begin{tabular}{cccccc}
    \toprule
    Method & CIFAR10 & CIFAR100 & Purchase100&Texas100&Location \\ 
    \midrule
    $m^*_{\textrm{DPSGD}}$&0.001&0.005& 0.05&0.01&0.5\\ 
    $\epsilon_{\textrm{DPSGD}}$&$\sim 10^9$&$ \sim 5\times10^7 $&$ \sim 10^5$&$\sim 10^6$&$\sim10^2$\\
    \midrule
    $m^*_{\textrm{DP-logits}}$& 0.01& 0.001& 0.001&0.0005&0.005 \\
    $\epsilon_\textrm{{DP-logits \lrnf}}$&$\sim 500$&$\sim 5000$&$ \sim 5000$&$\sim10^4$&$\sim10^3$\\
    $\epsilon_\textrm{{DP-logits \lrn}}$&$\sim 10^7$&$\sim 10^8$&$\sim 10^8$&$\sim 10^8$&$\sim 10^6$\\
    \bottomrule
    \end{tabular}
    \label{tab:optimalm}
\end{table*}

\begin{figure*}
    \centering
    \includegraphics[width=0.45\textwidth]{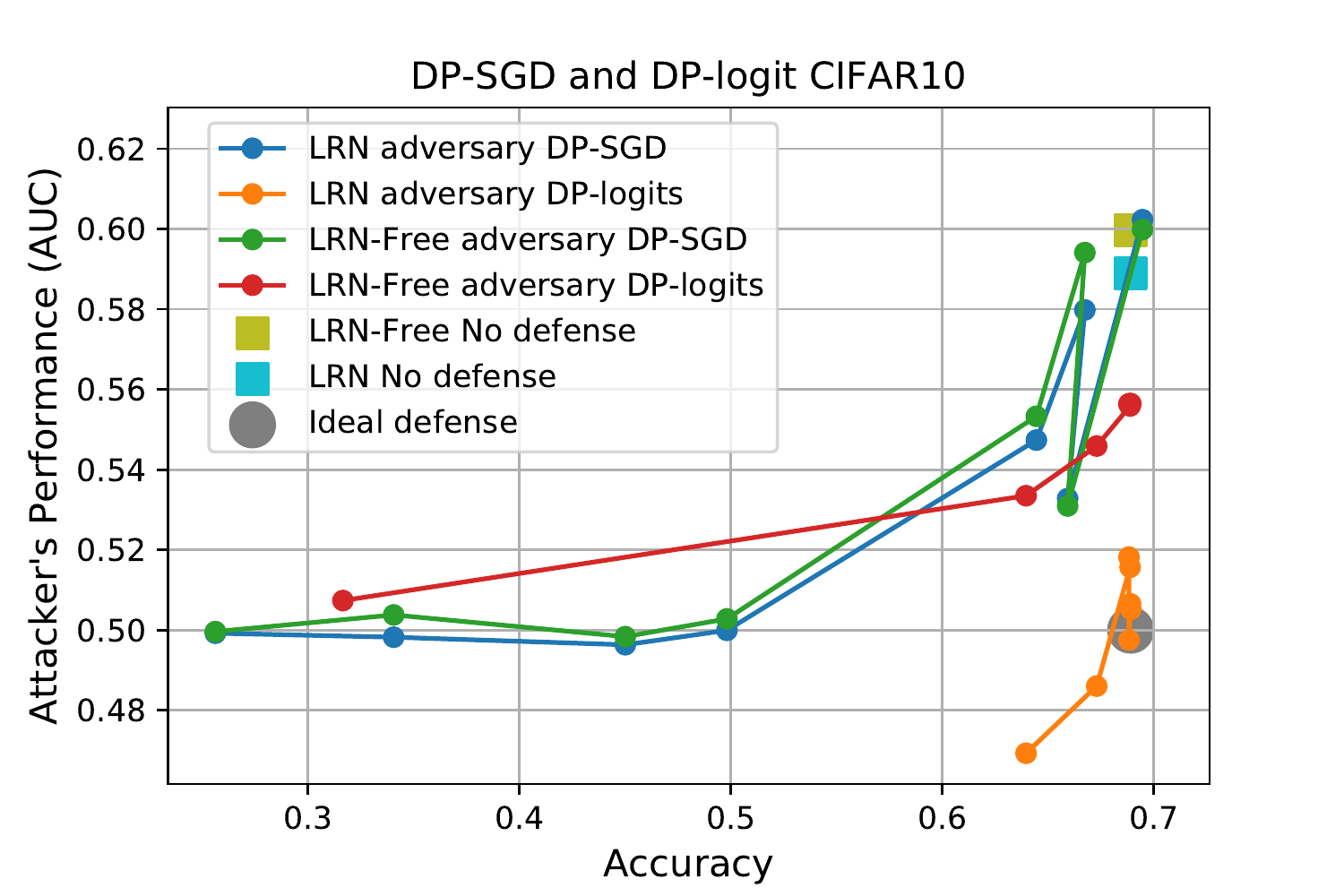}
    \includegraphics[width=0.45\textwidth]{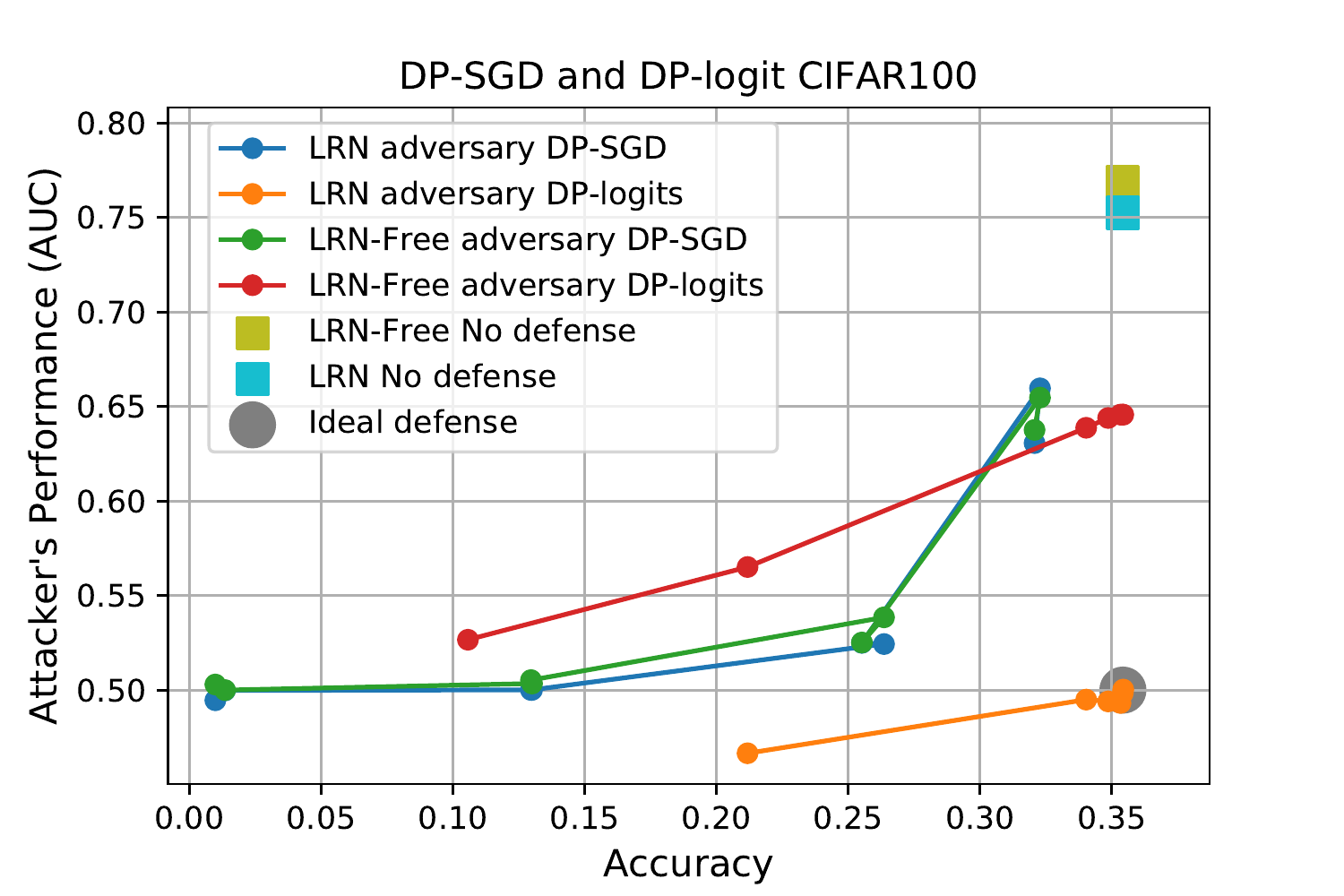}
    \includegraphics[width=0.45\textwidth]{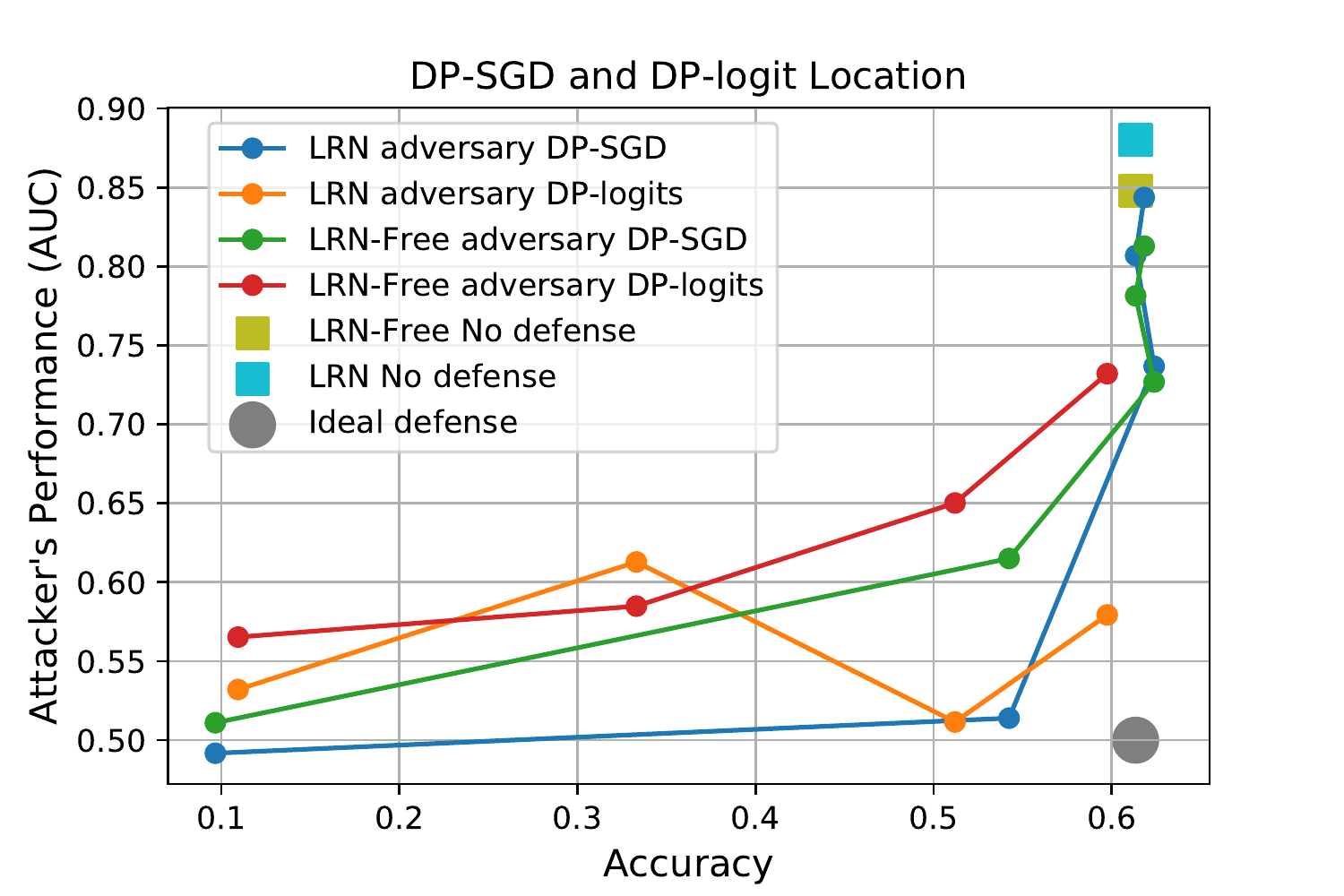}
    \includegraphics[width=0.45\textwidth]{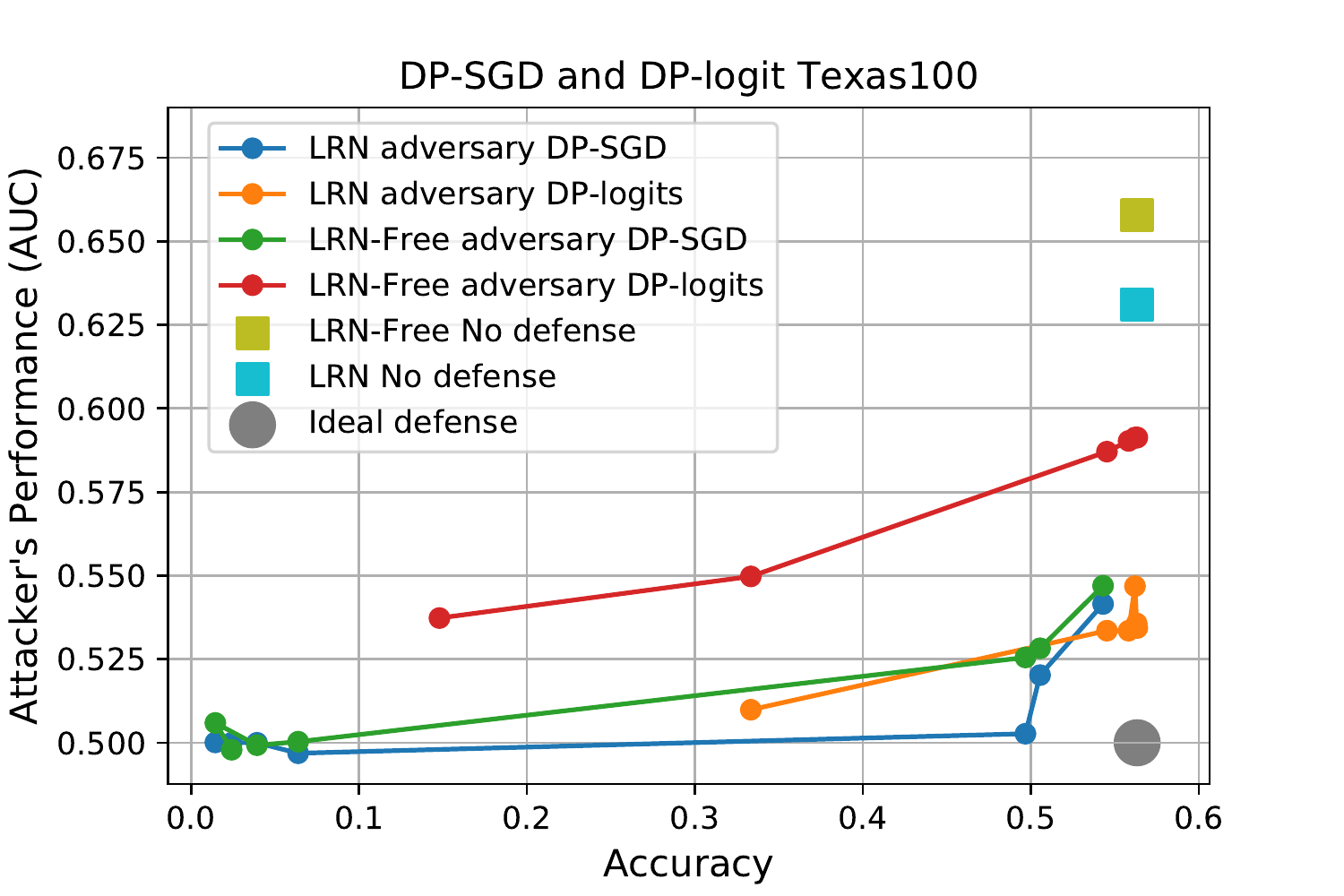}
    \includegraphics[width=0.45\textwidth]{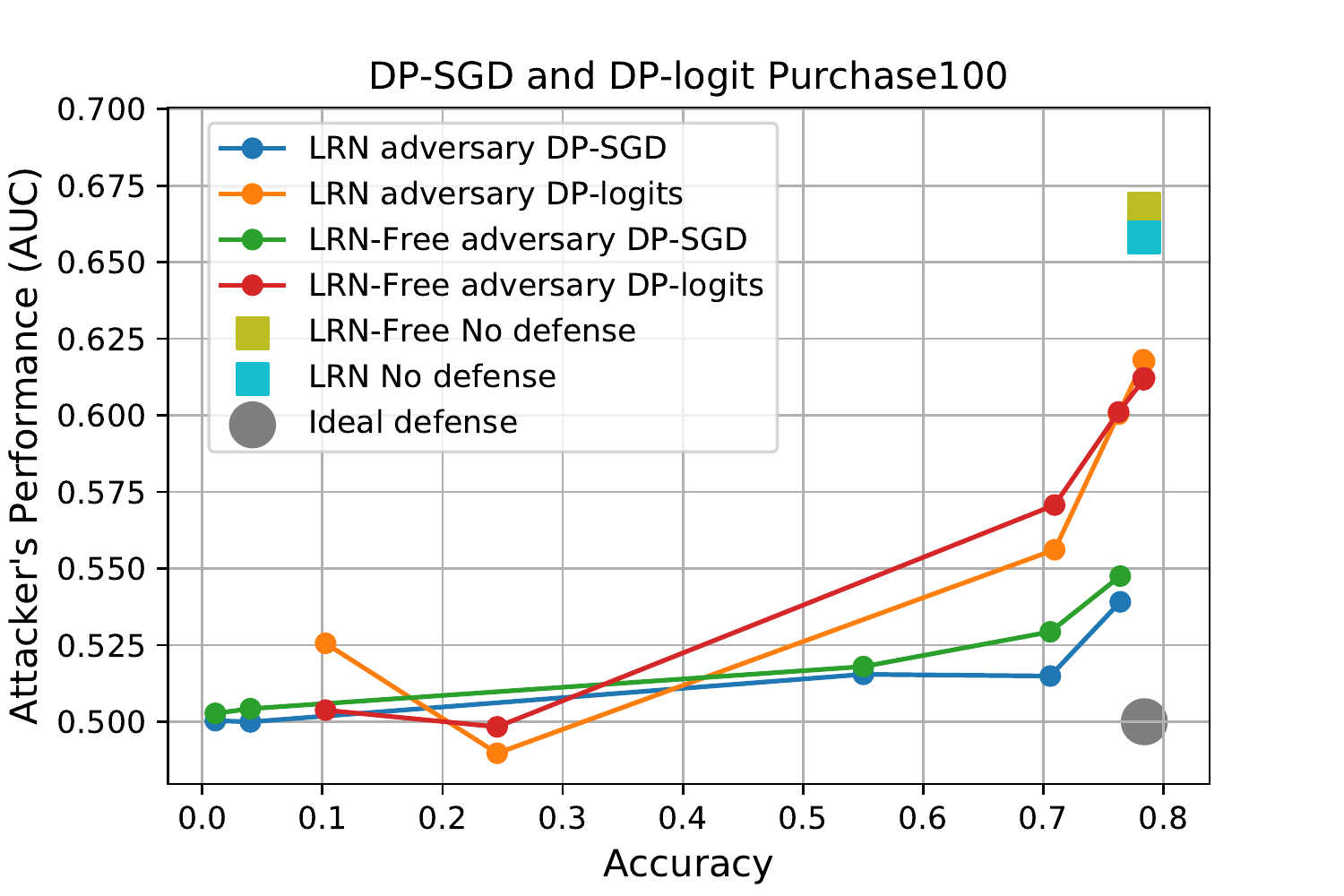}
    \caption{Results of application of DP-SGD and DP-logits for different noise levels. The accuracy of the vicitm model is shown on the x-axis. Each dot represents a noise level and they are connected with an increasing level of noise.}
    \label{fig:dpsgddplogits}
\end{figure*}

\myparagraph{Results and discussion}
Figure.~\ref{fig:dpsgddplogits} shows the trade-off of accuracy and privacy for each defense method and different levels of noise. Each color represents an adversary/defense combination and each dot represents a different noise level of the DP method. Dots are connected to each other in an increasing order of the value of noise. We observe that as the noise level rises the accuracy of the model as well as the performance of the adversary drop. What we are looking for in these plots are operating points near the grey "ideal defense" bubble. This indicates an area where the accuracy of the model does not suffer while the adversary is unable to perform a successful attack. 

Table~\ref{tab:optimalm} summarizes the (rounded) optimal values of noise multiplier $m$ and $\epsilon$ for DP-SGD and DP-Logits. We found these by inspecting the plots for the largest $m$ value (highest privacy) that falls within $~80\%$ of the initial accuracy.
We found that \lrn and \lrnf adversaries perform on the same level and even for the DP-Logits method with very different $\epsilon$ values, neither outperforms the other. This can be due to the fact that the \lrn is not designed to learn the noise structure of the DP-Logits. 

We also observe that the privacy budget for the DP-Logits method for \lrnf adversary is generally lower than the DP-SGD method. This was expected as the DP-SGD is designed to protect the whole model including the internal parameters and to achieve better test accuracy for the more complex victim models that are applied on more advanced datasets we need to allow a significant privacy budget. We see that these models are highly sensitive to noise and high values of noise that guarantees meaningful differential privacy bounds, completely destroys the utility of the model. For \lrn adversary repeated queries to the victim model degrades the privacy budget significantly, however, as mentioned we do not find any difference in performance of these adversaries.  

But perhaps the most curious finding is that despite the very large values of $\epsilon$, our DP methods offer good protection against membership inference attacks. This is most likely due to the fact that these adversaries utilize very little information from the models, for example no information from the internal parameters, and the budget of a DP method is defined to guarantee privacy under any adversarial technique and any auxiliary information. 

\subsection{Novel Sampling Attack and Defenses for It}
\label{sec:sampling-defenses}
With an understanding of how the conventional membership inference attacks perform in practice and how successful different DP mechanisms are to defend against them, we can start our experiments on the sampling attack. 
Similar to Section~\ref{sec:mi-defenses} we choose the datasets that show meaningful weaknesses towards adversaries.

The structure of the victim and shadow models are the same as the previous sections. 

Here also we decided to use two DP mechanisms for defense against the sampling adversary. We choose DP-SGD as a method that is applied during the training and randomized response that can be applied at prediction time (see Section~\ref{sec:dp} and~\ref{sec:dp-ml}).  

\myparagraph{Sampling adversary} For image datasets (i.e. CIFAR10/100) we choose pert$(x; p)=x + \mathcal{N}(0, p^2)$ which is an additive Gaussian noise to each pixel of the image in each channel. We chose  $p=\{i\times0.01|0\leq i\leq20, i\in\mathbb{N}\}$.

For binary datasets (Location, Texas100, Purchase100) pert$(x; p)= flip(x)$. So we flip the values of each dimension with a probability $p$ to the other value, that is, we flip 1 to 0 and 0 to 1. The steps of perturbation are chosen such that $p=\{i\times0.005|0\leq i\leq20, i\in\mathbb{N}\}$

For all of the datasets, we chose to generate $N=100$ perturbed samples for the attack. 

For steps 3 and 4 of the sampling adversary (see Section~\ref{sec:sampling}) We choose \lrnf as the conventional adversary. This is based on our findings from the previous section that showed no significant difference between the performance of \lrn and \lrnf. \lrnf is more versatile and reduces one extra step of the training an attack binary classifier for the sampling adversary. The structure of the \lrnf is the same as the previous sections. 

\myparagraph{DP-SGD} We use DP-SGD as the parameter perturbation mechanism against this adversary. We pick the optimum value of noise level from Table.~\ref{tab:optimalm} and run the sampling attack on the model trained with this noise level of DP-SGD.

\myparagraph{Randomized response} 
We choose randomized response (see Section~\ref{sec:dp}) on the returned labels, assuming that a trained model that follows \am protocol only allows adjusting of the labels and not the logits. Since we have more than two possible answers (labels), we modify the randomized response with a fair coin, in he following way: 

\textit{if tails:} reveal the returned label

\textit{if heads:} toss the fair coin again. If \textit{heads} reveal the returned label; otherwise uniformly at random choose one of the remaining classes.  

In this case, the privacy budget can be calculated as:
\begin{align}
    \label{eq:rrmod}
    e^{\epsilon} = \frac{\pr(\text{revealed}=\text{k}|\text{returned}=\text{k})}{\pr(\text{revealed}=\text{k}|\text{returned}=\text{k'})} &= \frac{0.75}{0.25/(C-1)}\nonumber\\
    &= 3(C-1)\nonumber
\end{align}
where $C$ is the total number of classes in the labels. So this mechanism is $(\ln(3C-3), 0)$ differentially private.

The expected accuracy of the model after using this method can be calculated as: 
\begin{align}
\textrm{accuracy}&=\frac{n_T}{n_T+n_F}\nonumber\\
\rightarrow \mathbb{E}[\textrm{accuracy}_{DP}] &= \frac{0.75*n_T}{n_T+n_F}+\frac{0.25/(C-1)* n_F}{n_T+n_F}\nonumber\\
&=0.75*accuracy + \frac{0.25}{C-1}(1-accuracy)\nonumber
\end{align}
where $n_T$ and $n_F$ indicate the number of points which correctly match the ground truth labels and the number of points which deviate from the ground truth labels, respectively. And $n_T+n_F$ is the total number of data points that the model was tested on. 
\begin{figure}
    \centering
    \includegraphics[width=0.45\textwidth]{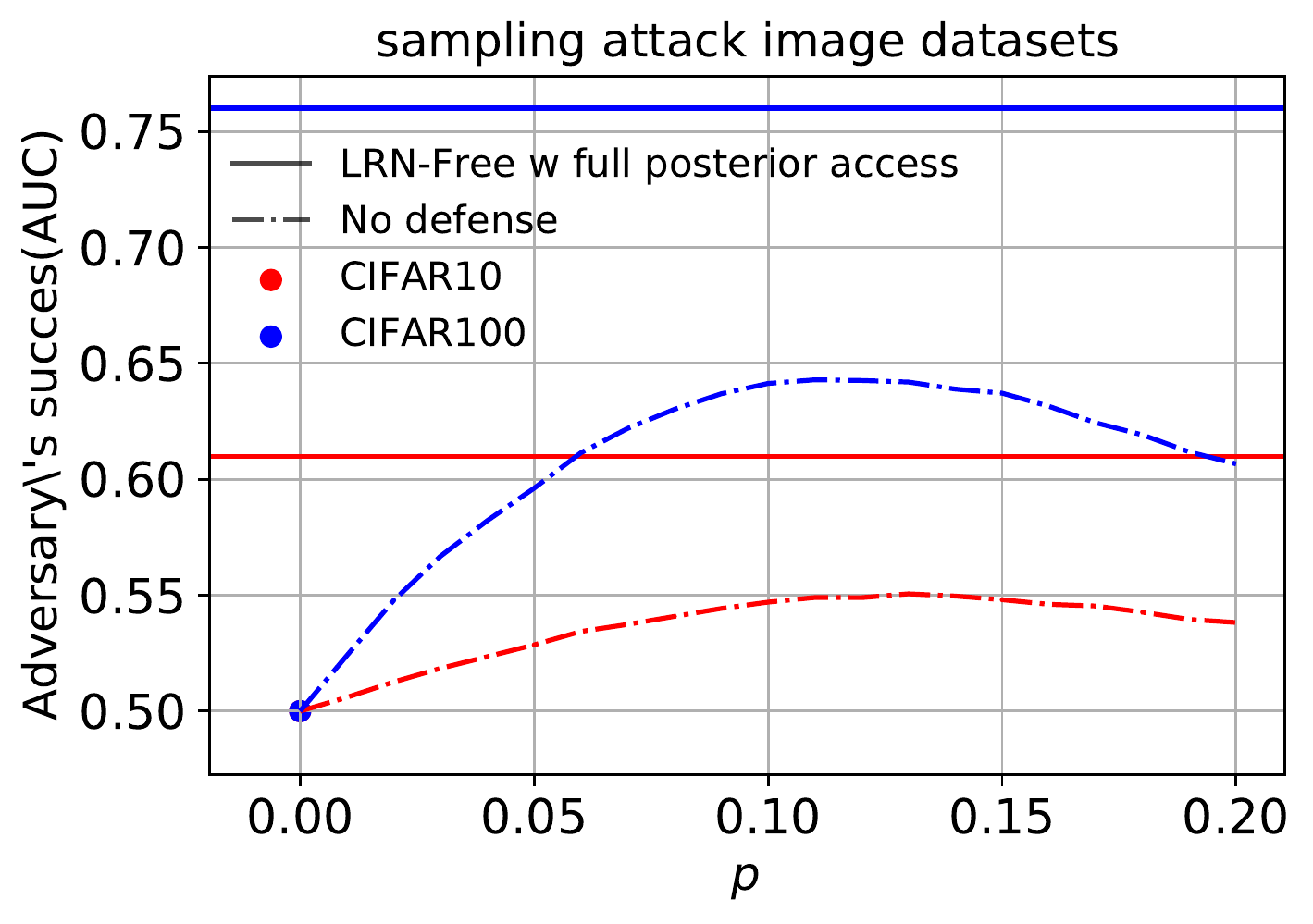}
    \includegraphics[width=0.45\textwidth]{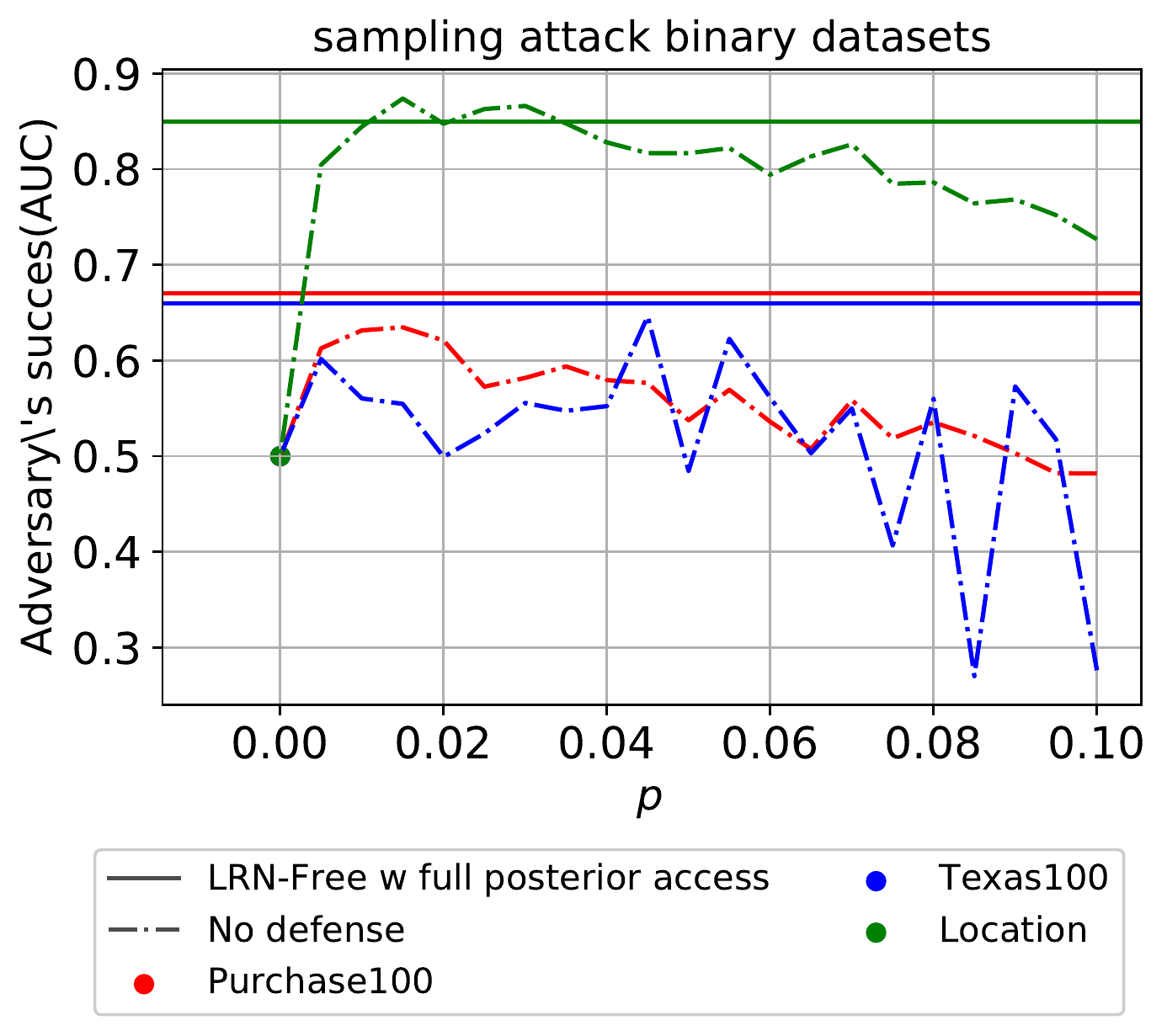}
    \caption{Sampling attack versus the \lrn adversary with full posterior access for binary and image datasets. On x-axis the perturbation scale is shown }
    \label{fig:sampling-both}
\end{figure}
\myparagraph{Results and discussion}
In Figure.~\ref{fig:sampling-both} we show the performance of the sampling adversary for different perturbation levels and compare it to the horizontal baseline of the \lrn adversary when the full posterior access is allowed. 

We observe that the sampling adversary is able to gain up to $50\%$ of its initial performance for image datasets and $100\%$ of the performance for binary datasets. The best result is achieved for Location dataset and this could correlate with the fact that location has the fewest number of features among all datasets and the perturbations around each data point are more meaningful and building the histograms over the labels is more representative of the true posteriors.

Table~\ref{tab:optimalp} includes the best perturbation scales $p$ with the highest AUC of adversary, found for each dataset. We see that this value is very small and particularly in the case of the binary datasets a $p<0.015$ works best. After this range, further perturbation results in random and noisy behavior of the sampling attack. 

Next, we pick $p^*$ and attack the victim model. Table~\ref{tab:samplingdefenses} shows the results for the \lrn with access to posteriors, the sampling attack AUC and sampling attack when DP-SGD and randomized response are used. Both DP-SGD and randomized response mitigate the risks of such attacks, however DP-SGD seems to protect better and the attacker's performance drops to almost chance level, for most datasets. 

\begin{table}[]
    \centering
    \caption{Best $p$ value for sampling attack for each dataset}
    \scalebox{0.85}{
    \begin{tabular}{cccccc}
    \toprule
    &CIFAR10&CIFAR100&Purchase100&Texas100&Location\\
    \midrule
    $p^*$&0.13&0.11&0.015&0.005&0.015\\
    \bottomrule
    \end{tabular}
    }%
    \label{tab:optimalp}
\end{table}

\begin{table}[]
 \caption{Sampling attack with the best $p$ value on the victim model.}
    \centering
\begin{tabular}{ccccc}
    \toprule
     Dataset &w/ access& sampling & DP-SGD & RR \\
     \midrule 
     CIFAR10&0.59 & 0.55&0.51&0.53\\
     CIFAR100&0.78& 0.66&0.51&0.61\\
     Purchase100&0.69& 0.67& 0.52&0.57\\
     Texas100&0.64& 0.63& 0.51&0.59\\
     Location&0.89& 0.89&0.61&0.8\\
     \bottomrule
\end{tabular}
    \label{tab:samplingdefenses}
\end{table}

We then study the effect of sampling number $N$ on the performance of the adversary on the victim model. We choose $N\in\{10, 100, 1000\}$ and calculate the AUC of the adversary for the optimum $p$ values from Table~\ref{tab:optimalp}. The results are shown in Table~\ref{tab:samplingnumbers}. As expected, higher number of queries improve the attack. This is because according to Equation.~\ref{eq:montecarlo} our estimate for the posterior limits to the true posterior when $n\rightarrow\inf$. It is important to mention that querying a victim model multiple times is not in general a desirable action. A victim model can detect and block multiple queries. We also observe that the increase in performance is more noticeable between $N=10-100$ than $N=100-1000$, so for these datasets $N=100$ is an acceptable and safe value.  

\begin{table}[]
 \caption{Effect of number of samples on attacks.}
    \centering
\begin{tabular}{ccccc}
    \toprule 
     Dataset &w/ access& N=10& N=100& N=1000 \\
     \midrule 
     CIFAR10&0.59 & 0.52&0.55&0.56\\
     CIFAR100&0.78& 0.60&0.66&0.68\\
     Purchase100&0.69& 0.57& 0.67& 0.68\\
     Texas100&0.64& 0.54& 0.63&0.63\\
     Location&0.89& 0.80&0.89&0.89\\
     \bottomrule 
\end{tabular}
    \label{tab:samplingnumbers}
\end{table}

At last, we also show the AUC values of the attack when the optimum $p$ of another dataset is used to attack. Table~\ref{tab:diffp} shows the AUC of adversary on the victim model for different $p^*$ of datasets. We transfer the best perturbation scale among image datasets and binary datasets, separately. These results show us that an acceptable attack performance can be achieved even when the adversary trains the shadow model for a different dataset. With this strategy, the attacker is able to train the shadow model once on a similar type of dataset and carry out attacks on other data and save on the training time.  

\begin{table}[]
 \caption{Transferring the best $p$ between datasets}
    \centering
\begin{tabular}{ccccc}
    \toprule
     $p$ &0.11&0.13&0.005&0.015\\
     \midrule
     CIFAR10&0.55&0.54&&\\
     CIFAR100&0.64&0.63&&\\
     Purchase100&&& 0.65&0.67\\
     Texas100&&& 0.63&0.54\\
     Location&&&0.81&0.89\\
     \bottomrule 

\end{tabular}
    \label{tab:diffp}
\end{table}

%% file: conclusion.tex
\section{Discussion and Conclusion}
\label{sec:conclusion}
By evaluating membership inference attacks over a large scope of different dataset, we highlight issues with the rapidly developing research thread of membership inference. Often attack performance is assessed with different type of models, while such performance cannot be seen in isolation of data and training procedure. We urgently need more transparency in reporting membership attack performance in order to be really in a position to compare and measure progress in this area. While a large fraction of the work is focused on attacks, we also provide defense evaluation. We investigate ``standard'' DP-SGD, but also look at effective ``post-hoc'' defenses, that might be a lot more practical (e.g. faced with ``legacy'' models) as they can operate on already trained models and provide strong protection.

Most notably, we investigate the \textit{argmax} defense, which previously was thought to prevent membership inference attacks. While for previous attacks this is true, we show a new sampling attack that attracts by repeated query of the model surrogate information of the model that is able to recover a large fraction of the attack performance. In turn, we also present a modification of the randomized response defense, that is in part capable of mitigating the new attack vector.

At time of publication, we will make all our attacks and defenses publicly available -- which constitute the largest assembly of techniques in the scope of membership inference attacks that we know of. We believe that this is a contribution that the community will also benefit from. We will also publish our victim models and encourage to re-use these as a point of reference, but advocate transparency and advise to publish more details on training procedures in future work as otherwise comparable results and ultimately the scientific progress in this area is compromised.